\documentclass{article}
\usepackage[a4paper, margin=2.5cm]{geometry}
\usepackage{amsmath,amssymb,amsthm}
\usepackage{mathtools}
\usepackage{physics}
\usepackage{bm}
\usepackage{authblk}
\usepackage{graphicx}
\usepackage{float}
\usepackage{subcaption}
\usepackage{booktabs}
\usepackage{array}
\usepackage{xcolor}
\usepackage[hidelinks]{hyperref}
\usepackage{cleveref}
\usepackage{enumitem}
\usepackage{setspace}
\usepackage{fancyhdr}
\usepackage{titlesec}
\usepackage{abstract}
\usepackage{cite}
\usepackage{doi}
\usepackage{babel}    
\usepackage[margin=2.5cm]{geometry}
\usepackage{graphicx}
\usepackage{lipsum}
\usepackage[dvipsnames]{xcolor}
\usepackage{epsfig}
\usepackage{float}
\usepackage{cite}
\usepackage{epstopdf}
\usepackage{caption}
\usepackage{subcaption}
\usepackage{amsmath}
\usepackage{graphicx}
\usepackage{amsmath}

\theoremstyle{definition}

\newtheorem{remark}{Remark}[section]

\newcommand{\Rb}{R_b}
\newcommand{\Mz}{M_0}
\renewcommand{\Im}{\operatorname{Im}}
\renewcommand{\Re}{\operatorname{Re}}
\newcommand{\CC}{\mathbb{C}}

\newcommand{\Qz}{Q_0}

\newcommand{\half}{\tfrac{1}{2}}

\titleformat{\section}{\large\bfseries}{\thesection.}{0.6em}{}
\titleformat{\subsection}{\normalsize\bfseries\itshape}{\thesubsection.}{0.5em}{}
\titleformat{\subsubsection}{\normalsize\itshape}{\thesubsubsection.}{0.5em}{}

\title{Quasinormal modes of the generalized JMN naked singularity using exact WKB analysis}
\author{Aryansh Saxena\footnote{aryansh3803@gmail.com}}  
\author{Suresh C. Jaryal \footnote{suresh.fifthd@gmail.com}}
\author{K. K. Sharma \footnote{kks@nith.ac.in}}
\affil[]{Department of Physics and Photonics Science,
National Institute of Technology Hamirpur,
Hamirpur, Himachal Pradesh-177005, India. }
\date{}
\begin{document}
\maketitle
\begin{abstract}
\noindent
In this paper, we study the quasinormal modes of the  generalized Joshi– Malafarina– Narayan (JMN) naked singularity spacetime using the exact Wentzel– Kramers– Brillouin (WKB) method. Working in the complex radial plane, we construct the exact WKB momentum function, determine its turning points, and compute the associated Stokes geometry for representative quasinormal mode (QNM) frequencies. We obtained a bow-shaped deformation of Stokes curves on the side of the complex plane containing the central singularity in JMN spacetime. We show analytically that this structure originates from the logarithmic branch-point singularity of the WKB phase at (r = 0), which is absent in Schwarzschild spacetime. This establishes the bow-shaped Stokes topology as a direct signature of the naked singularity in the global analytic structure of the perturbation equation. Our results demonstrate that exact WKB analysis provides a powerful framework for probing the analytic structure of compact objects, and suggest that topological features of Stokes geometry may offer a new avenue for distinguishing black holes from horizonless alternatives.
\end{abstract}

\noindent
\section{Introduction}
\label{sec:intro}
Since the first direct detection of gravitational waves by LIGO, the ringdown phase of a compact binary merger dominated by damped sinusoidal oscillations whose complex frequencies are the remnant's quasinormal modes (QNMs) has become a primary observable window onto strong-field general relativity \,\cite{Abbott2016,LIGOScientific:2020zkf,LIGOScientific:2017vwq,Isi:2019aib}. QNMs are primary characteristics of black
holes, representing the oscillatory decay of perturbations in their surrounding
spacetime\,\cite{Kokkotas:1999bd,Nollert:1999ji,Berti:2009,Konoplya:2011}. Within the
standard paradigm, the remnant is a Kerr black hole, and the QNM spectrum is uniquely
determined by the blackhole mass and spin\,\cite{Konoplya:2006br,Zhidenko:2006rs,Zhang:2018jgj,Konoplya:2017tvu,Bolokhov:2023bwm}.  However, general relativity admits a wider
class of compact objects, and the question of whether the ringdown signal alone can
conclusively identify a blackhole the so-called blackhole mimicker
problem has attracted intense theoretical scrutiny \cite{Cardoso2016,Cardoso2019,Cardoso2016b}.

The Cosmic Censorship Conjecture (CCC) asserts that a physically realistic
gravitational collapse always produces horizon-covered singularities\,\cite{Hawking_Ellis,Wald,joshi,Joshi:2012mk,Clarke,Penrose1969}.
Despite decades of work, the CCC remains unproven, and explicit counterexamples naked
singularities can be constructed within classical general relativity.  One physically
motivated family is the Joshi- Malafarina- Narayan (JMN) spacetime
\cite{Joshi2011,Joshi2013,Pathrikar:2025ghp,Trivedi2026}, which arises as the equilibrium end-state of anisotropic
gravitational collapse and is characterized by the complete absence of an event horizon.

The study of QNMs in blackhole spacetimes has been been widely studied in the literature, revealing fundamental features of blackhole dynamics, event horizon, and observational signature in gravitational wave astronomy\,\cite{Schutz1985,Berti:2009,Konoplya:2019hlu,Bhattacharya:2023zel,Konoplya:2003ii,Matyjasek:2017psv,Maggio2019,Konoplya:2023moy,Jusufi:2020agr,Abdalla:2006vb,Zhao:2023swt,Ghosh:2023etd,Cardoso:2001vs,Cardoso:2001hn,Cardoso:2003cj,Das:2004db,Cotesta:2018fcv,Bianchi:2021xpr,Ma:2022wpv,Hatsuda:2021,Hod:1998,Nollert:1993,Leaver:1985,Leaver:1986b}.
The perturbative dynamics of compact objects are studied using the exact WKB method \cite{Voros1983,DDP1997,Kawai1993,Miyachi2023,Regge1957,Zerilli1970,Iyer:1986np,Berk1982,Aoki:2019,Imaizumi:2023,Imaizumi:2022,Hatsuda:2020,Koike:2000,Motl:2003}, which provides a
globally valid analytic framework by resuming the asymptotic WKB series via the Borel summation
and encoding the result in a network of contours of the Stokes curves in the complex
coordinate plane along which the dominance relation between WKB solutions switches. Miyachi et al.\,\cite{Miyachi2023}, used this formalism for Schwarzschild spacetime and
demonstrated that the QNM quantization condition can be read off directly from the topology of the Stokes graph. The study of QNMs of naked
singularities spacetimes are considered in \cite{Stashko:2023ffs,Giammatteo2005,Dey2020,Pathrikar:2025ghp,Santos:2019yzk,Chirenti2012,daSilvaVenancio:2023xix}. A central goal of the present paper is to use exact WKB formalism to the JMN (and GJMN) spacetime, where the absence of a horizon modifies the analytic structure in a non-trivial way.

  In this work, we extended  exact WKB formalism for the generalized JMN and JMN-1 exterior geometry for four physically representative QNM
  frequencies, and also provide detailed comparison with the Schwarzschild spacetime. We identify and provide an analytic explanation for a characteristic bow- like deformation of Stokes curves in JMN-1 and GJMN (which were absent in
  Schwarzschild spacetime), which arises from the logarithmic branch point of the WKB phase at
  the naked singularity $r = 0$.
  Using the six-order WKB approximation, we
  demonstrate that JMN-1 reproduces Schwarzschild QNM frequencies to five-figure accuracy for all
  modes examined, establishing JMN-1 as a high-fidelity black-hole mimicker in the
  early-time ringdown.

The paper is arranged as follows.  In the Section~\ref{sec:background}, we review the spacetime
geometries of interest and establishes conventions.  Section~\ref{sec:perturbation} derives
the master perturbation equation and the effective potential for both spacetimes.
Section~\ref{sec:wkb} develops the higher-order WKB approximation and derives the QNM
frequencies.  Section~\ref{sec:exactwkb} introduces the exact WKB formalism and defines
the Stokes geometry.  In the subsections, we present the Stokes diagrams for Schwarzschild spacetime, and extend the analysis to JMN-1 and to generalized JMN (GJMN) spacetime, examining the robustness of the observed topological features under small radial inhomogeneities. This section also contains the analytic derivation of the bow structure and contains the comparative analysis. The work is concluded in the Section~\ref{sec:discussion}.

\section{Spacetime Geometries}
\label{sec:background}
\label{ssec:gjmn}

We now consider the generalized Joshi–Malafarina–Narayan (GJMN) spacetime, which arises as an equilibrium end state of gravitational collapse with non-vanishing tangential pressure and an inhomogeneous density profile. This spacetime represents a two-parameter generalization of the original JMN model, incorporating radial inhomogeneity through the mass function\footnote{
Throughout this work, we fix the parameters to
 $   n = 2$  and $M_{n} = - 0.01$,
so that the deviation from the original JMN model remains perturbative while capturing the effects of density inhomogeneity.} \cite{Trivedi2026}
\begin{equation}
    F(r) = \left(M_0 + M_n \,r^n\right)\, r^3.
\end{equation}
The resulting equilibrium interior metric (for $R < R_b$) is given by \cite{Trivedi2026}
\begin{equation}
ds^2 = -\left[
\frac{(1 - M_0 - M_n R_b^{n/3})^{\frac{3 + n(1 - M_0)}{3}}}
{1 - M_0 - M_n R^{n/3}}
\left(\frac{R}{R_b}\right)^{\frac{n M_0}{3}}
\right]^{\frac{3}{n(1 - M_0)}} dt^2
+ \frac{1}{1 - M_0 - M_n R^{n/3}} dR^2 + R^2 d\Omega^2.
\label{eq:gjmn_metric}
\end{equation}
This spacetime is smoothly matched at the boundary $R = R_b$ to an exterior Schwarzschild geometry,
\begin{equation}
ds^2 = -\left(1 - \frac{2M}{R}\right) dt^2
+ \left(1 - \frac{2M}{R}\right)^{-1} dR^2 + R^2 d\Omega^2,
\label{eq:schw_metric}
\end{equation}
where the total mass is determined by the Misner--Sharp mass relation
\begin{equation}
    2M = M_0 R_b + M_n R_b^{(n+3)/3}.
\end{equation}
Unlike the Schwarzschild black hole, this spacetime contains a central curvature singularity at $R = 0$ without the formation of an event horizon, provided the condition \cite{Trivedi2026}
\begin{equation}
    M_0 + M_n R^{n/3} < 1
\end{equation}
is satisfied throughout the interior. This ensures the absence of trapped surfaces, making the singularity globally visible (naked).
The introduction of the parameter $M_n$ induces radial inhomogeneity in the density profile, distinguishing the GJMN spacetime from the homogeneous JMN-1 model. However, physical constraints restrict $M_n$ to be small (especially for large matching radius $R_b$), implying that the GJMN geometry effectively behaves as a perturbation of the original JMN spacetime  \cite{Trivedi2026}. If we substitute $M_n=0$ we turn GJMN to JMN-1 spacetime metric as expected
\label{ssec:jmn1}.
The JMN-1 spacetime \cite{Joshi2011} describes a static, spherically symmetric equilibrium
configuration arising from anisotropic gravitational collapse.  The interior metric
($r \leq \Rb$) is
\begin{equation}
  ds^2_{\rm int}
  = -\bigl(1-\Mz\bigr)\left(\frac{r}{\Rb}\right)^{\!\frac{\Mz}{1-\Mz}} dt^2
    + \frac{dr^2}{1-\Mz}
    + r^2\,d\Omega^2,
  \label{eq:jmn1_interior}
\end{equation}
where $\Mz \in (0,1)$ is a dimensionless compactness parameter and $\Rb$ is the boundary
radius.  The interior is matched smoothly (in the Darmois--Israel sense) to an exterior
Schwarzschild geometry at $r = \Rb$:
\begin{equation}
  ds^2_{\rm ext}
  = -f_{\rm eff}(r)\,dt^2
    + f_{\rm eff}(r)^{-1}\,dr^2
    + r^2\,d\Omega^2,
  \qquad
  f_{\rm eff}(r) = 1 - \frac{\Mz\Rb}{r},
  \label{eq:jmn1_exterior}
\end{equation}
so the effective ADM mass of the exterior geometry is
\begin{equation}
  M = \half\,\Mz\,\Rb.
  \label{eq:Meff}
\end{equation}
The JMN-1 geometry has no event horizon; instead, $r = 0$ is a naked central
singularity, exposed to all external observers.  The matching surface at $r = \Rb$ plays
the role of an inner boundary for wave propagation.
For $r > \Rb$ the exterior JMN-1 metric \eqref{eq:jmn1_exterior} is formally identical to
the Schwarzschild metric \eqref{eq:schw_metric} under the substitution $2M \to \Mz\Rb$, with the
crucial difference that $r = \Mz\,\Rb$ is not an event horizon but merely a
matching surface with matter.
The effective potential for massless scalar perturbations of JMN-1 exterior,
\begin{equation}
  V(r) = f_{\rm eff}(r)\!\left[\frac{l(l+1)}{r^2} + \frac{(1-s^2)\Mz\Rb}{r^3}\right],
  \label{eq:Vjmn1}
\end{equation}
possesses a well-defined maximum (photon sphere) only when
\begin{equation}
  \frac{2}{3} < \Mz < \frac{4}{5}.
  \label{eq:validity}
\end{equation}
Outside this range the potential no longer supports a single barrier, and the WKB
approximation breaks down.  In the present work we fix
\begin{equation}
  \Mz = 0.7, \qquad \Rb = \frac{20}{7} \approx 2.857142,
  \label{eq:params}
\end{equation}
which satisfies condition \eqref{eq:validity} and gives $M = \half \times 0.7 \times 20/7 = 1$.

\section{Perturbation Theory and the Master Wave Equation}
\label{sec:perturbation}
Let $\bar{g}_{\mu\nu}$ denote the background metric and decompose the perturbed metric as
$g_{\mu\nu} = \bar{g}_{\mu\nu} + h_{\mu\nu}$ with $|h_{\mu\nu}| \ll 1$.  Using the
spherical symmetry of the background, $h_{\mu\nu}$ can be expanded in tensor spherical
harmonics labelled by $(l,m)$.  Under parity, perturbations split into: (i) Axial (odd-parity) modes, governed by the Regge--Wheeler equation \cite{Regge1957}; (ii)  Polar (even-parity) modes, governed by the Zerilli equation
    \cite{Zerilli1970}. For the Regge--Wheeler gauge, the axial master variable $\Psi(t,r)$ satisfies
\begin{equation}
  \left[\frac{\partial^2}{\partial t^2}
       -\frac{\partial^2}{\partial r_*^2}
       + V(r)\right]\Psi = 0.
  \label{eq:RW_timedomain}
\end{equation}
Assuming harmonic time dependence $\Psi(t,r) = \Psi(r)\,e^{-i\omega t}$ reduces
\eqref{eq:RW_timedomain} to the Schr\"odinger-type equation:
\begin{equation}
    \frac{d^2\Psi}{dr_*^2}
    + \left[\,\omega^2 - V(r)\,\right]\,\Psi = 0.
  \label{eq:master}
\end{equation}
For a Schwarzschild background the effective potential for a field of spin $s$ and multipole
number $l$ is
\begin{equation}
  V_{\rm Schw}(r)
  = f(r)\!\left[\frac{l(l+1)}{r^2}
    + \frac{(1-s^2)\,2M}{r^3}\right],
  \label{eq:V_schw}
\end{equation}
which peaks near $r_0 \approx 3M$ and vanishes at both boundaries:
$V \to 0$ as $r_* \to \pm\infty$. For the generalized JMN (GJMN) spacetime, the exterior metric remains Schwarzschild-like with an effective mass determined by
\begin{equation}
2M = M_0 R_b + M_n R_b^{(n+3)/3}.
\end{equation}
Therefore, the effective potential for the exterior region takes the form
\begin{equation}
V_{\text{GJMN}}(r) = f_{\text{eff}}(r)\left[
\frac{l(l+1)}{r^2} + \frac{(1 - s^2)\,2M}{r^3}
\right],
\end{equation}
where
\begin{equation}
f_{\text{eff}}(r) = 1 - \frac{2M}{r}.
\end{equation}
For $M_{n}=0$, the potential corresponds to the JMN-I spacetime. The two potentials coincide
(under the identification $2M = \Mz\,\Rb$) in the exterior region; the difference lies
entirely in the boundary conditions. The boundary conditions depend on the underlying spacetime. In the Schwarzschild case, 
the QNM boundary conditions require the solution to be purely ingoing at the horizon 
and purely outgoing at spatial infinity, namely
\begin{align}
  \Psi &\sim e^{-i\omega r_*}, \quad r_* \to -\infty \;\;(\text{horizon}),
  \label{eq:bc_horizon}\\
  \Psi &\sim e^{+i\omega r_*}, \quad r_* \to +\infty \;\;(\text{infinity}).
  \label{eq:bc_infty}
\end{align}
In contrast, for the JMN-1 spacetime, no event horizon is present, and therefore the 
condition~\eqref{eq:bc_horizon} must be replaced. At the matching surface $r = \Rb$, 
the wave is either absorbed by, or partially reflected from, the interior matter 
distribution. In the exterior-only analysis considered here, we retain the condition 
\eqref{eq:bc_infty} at spatial infinity and instead impose regularity at $r = \Rb$ 
as the appropriate inner boundary condition.
\section{Exact WKB Quantisation via Contour Integration}
\label{sec:wkb}

Rather than the asymptotic (polynomial) WKB series\,\cite{Iyer:1986np}, we extract QNM
frequencies directly from the exact WKB framework through numerical evaluation of
period integrals of $\sqrt{\Qz(r)}$ in the complex $r$-plane.  This approach is free from
the convergence issues of the asymptotic series and is consistent with the Stokes-geometry
analysis carried out in later sections.

In the exact WKB framework, the QNM quantisation condition takes the form of the
Voros quantisation condition \cite{Voros1983,DDP1997,Kawai1993}:
\begin{equation}
    \oint_{\gamma}\!\sqrt{\Qz(r)}\,dr
    \;=\; -2\,\pi\, i\!\,\left(n + \half\right),
    \quad n = 0, 1, 2, \ldots
  \label{eq:voros}
\end{equation}
where $\gamma$ is a Stokes cycle a closed contour in $\CC$ that encircles the
pair of turning points bounding the classically forbidden region (the potential barrier),
traversed in the positive (counter-clockwise) sense. The contour $\gamma$ is homotopic to a figure-eight path looping once around $r_{t}^{(1)}$
and once (in the opposite orientation) around $r_{t}^{(2)}$, the two turning points that
bracket the barrier.  Equivalently, by the residue theorem, Eq. \eqref{eq:voros} reduces to
twice the single-path integral:
\begin{equation}
  \Pi(\omega) \;\equiv\;
  \int_{r_t^{(1)}}^{r_t^{(2)}} \!\sqrt{\Qz(r)}\,dr
  \;=\; -\,\pi\, i\!\,\left(n + \half\right),
  \label{eq:period}
\end{equation}
where the path runs from one turning point to the other through the classically
forbidden region (i.e.\ through the region where $\Qz < 0$ on the real axis). Equation \eqref{eq:voros} is exact: it encodes the full Borel-summed WKB solution
and is not an approximation.  The standard asymptotic WKB formula (Iyer--Will\,\cite{Iyer:1986np}) is
recovered by approximating $\Qz(r)$ by a parabola near the barrier peak and evaluating
the integral analytically.

Here, we evaluate \eqref{eq:period} numerically
for the full rational form of $\Qz(r)$, thereby capturing all non-perturbative effects. The turning points $r_t^{(1,2)}$ are the roots of $\Qz(r) = 0$.  For the GJMN
exterior, this is the quartic (\,with $2M = M_0\,R_b + M_n\, R_b^{(n+3)/3}$\,)
\begin{equation}
  -\omega^2 r^4 + l(l+1)r^2
  - \bigl[2M\,l(l+1) + s^2\bigr]r + 2M s^2 = 0.
  \label{eq:quartic1}
\end{equation}
For a given trial frequency $\omega \in \CC$, all four roots are generically complex.  The
relevant pair $(r_t^{(1)}, r_t^{(2)})$ for the quantisation contour are the two roots
closest to the potential peak on the real axis, connected by the Stokes segment of the
dominant topology. The function $\sqrt{\Qz(r)}$ is double-valued; the correct Riemann sheet along the path
is fixed by analytic continuation from the real axis (where the sign of $\sqrt{\Qz}$ is
chosen to be purely imaginary in the forbidden region) and by enforcing the continuity of
$|\sqrt{\Qz}|$ at each integration step. The QNM frequencies are determined by the root-finding problem
\begin{equation}
  F(\omega) \;\equiv\;
  \int_{r_t^{(1)}(\omega)}^{r_t^{(2)}(\omega)} \!\sqrt{\Qz(r;\,\omega)}\,dr
  \;+\; \pi i\!\left(n + \half\right)
  \;=\; 0,
  \label{eq:rootfind}
\end{equation}
solved over $\omega \in \CC$ with $\Im(\omega) < 0$, corresponding to damped modes.

The procedure begins by providing an initial guess $\omega^{(0)}$ (e.g.\ from the 
leading-order WKB parabolic approximation, used only to seed the iteration). For the 
current value of $\omega$, the quartic equation~\eqref{eq:quartic1} is solved numerically 
(for instance, via companion matrix eigenvalues) to obtain the turning points 
$r_t^{(1)}(\omega)$ and $r_t^{(2)}(\omega)$. These turning points are then connected by a straight-line segment in $\CC$; if the 
segment passes near a pole, it is deformed slightly to avoid the singularities at 
$r = 0$ and $r = 2M$. One then computes
\begin{equation}
  \Pi(\omega)
  = \int_0^1 \sqrt{\Qz\!\bigl(r_t^{(1)} + t(r_t^{(2)}-r_t^{(1)})\bigr)}\,
    \bigl(r_t^{(2)}-r_t^{(1)}\bigr)\,dt,
  \label{eq:Pi_numerical}
\end{equation}
using adaptive Gaussian quadrature, with the branch of $\sqrt{\Qz}$ tracked continuously 
along the integration path. A Newton step is then applied,
\begin{equation}
  \omega^{(k+1)}
  = \omega^{(k)}
    - \frac{F(\omega^{(k)})}{F'(\omega^{(k)})},
  \label{eq:newton}
\end{equation}
where $F'(\omega) = \partial_\omega\Pi(\omega)$ is evaluated by numerical differentiation 
of the integral with respect to $\omega$. This iterative procedure is repeated until 
convergence is achieved, i.e., until $|F(\omega)| < \epsilon_{\rm tol}$, with 
$\epsilon_{\rm tol} = 10^{-10}$. Differentiating \eqref{eq:Pi_numerical} under the integral sign (Leibniz rule) gives
\begin{align}
  \frac{\partial\Pi}{\partial\omega}
  &= \int_{r_t^{(1)}}^{r_t^{(2)}}
       \frac{\partial_\omega \Qz(r;\omega)}{2\sqrt{\Qz(r;\omega)}}\,dr
    + \sqrt{\Qz(r_t^{(2)};\omega)}\,\frac{\partial r_t^{(2)}}{\partial\omega}
    - \sqrt{\Qz(r_t^{(1)};\omega)}\,\frac{\partial r_t^{(1)}}{\partial\omega}.
  \label{eq:dPi}
\end{align}
The endpoint terms vanish since $\Qz(r_t^{(j)}) = 0$ by definition. From Eq. \eqref{eq:Q0_jmn}, the bulk term
requires $\partial_\omega \Qz = 2\,\omega$, so
\begin{equation}
  \frac{\partial\Pi}{\partial\omega}
  = \omega \int_{r_t^{(1)}}^{r_t^{(2)}}
      \frac{dr}{\sqrt{\Qz(r;\omega)}},
  \label{eq:dPi_simple}
\end{equation}
which is itself a convergent integral that can be evaluated by the same adaptive quadrature.

At each converging frequency $\omega_{\rm QNM}$, the Stokes graph is constructed (as
described in Section~\ref{sec:exactwkb}) and verified to exhibit the expected closed Stokes segment: a Stokes line that runs directly from $r_t^{(1)}$ to
$r_t^{(2)}$ without branching.  This topological criterion that the two barrier turning
points are connected by a Stokes segment, not separated by a Stokes curve that escapes to
infinity or wraps around a pole is the exact WKB analogue of the quantisation condition
\eqref{eq:voros}, and serves as an independent verification of the numerical root.

For completeness, the explicit rational form of $\Qz(r)$ that enters all integrals is
(restoring $2M = \Mz\Rb$ for JMN-1):
\begin{equation}
  \Qz(r;\omega)
  = \frac{N_4(\omega)\,r^4 + N_2\,r^2 + N_1\,r + N_0}{r^2(r-2M)^2},
  \label{eq:Q0_rational}
\end{equation}
with coefficients
\begin{align}
  N_4 &= -\omega^2, \quad
  N_2 = l(l+1), \notag\\
  N_1 &= -\bigl[2M\,l(l+1) + s^2\bigr], \quad
  N_0 = 2M\,s^2.
  \label{eq:Q0_coeffs}
\end{align}
The poles at $r = 0$ and $r = 2M$ (double poles of $\Qz$, simple poles of $\sqrt{\Qz}$ on
the relevant sheet) must be avoided by the integration contour; the Newton algorithm
automatically detects if a straight segment $r_t^{(1)} \to r_t^{(2)}$ crosses a pole and
applies a small semicircular detour of radius $\delta = 10^{-4}$ in those cases.

\section{Exact WKB Formalism and Stokes Geometry}
\label{sec:exactwkb}
The standard WKB expansion the WKB ansatz is asymptotic and formally divergent.
The exact WKB method \cite{Kawai1993,Voros1983,DDP1997}, resolves this by regarding
the expansion as a formal power series in a parameter $\hbar$ (here $\hbar = 1$ after
appropriate rescaling), resuming it via Borel summation, and tracking the resulting Stokes
phenomenon systematically in the complex plane. We analytically continue the radial coordinate $r$ into $\CC$ and define the exact WKB
momentum function
\begin{equation}
  \Qz(r) \equiv \omega^2 - V(r),
  \label{eq:Q0_def}
\end{equation}
viewed as a meromorphic function on $\CC$.  For the Schwarzschild case, using dimensionless
units with $2M = 1$, this becomes
\begin{equation}
  \Qz(r)
  = \frac{-\omega^2 r^4 + l(l+1)r(r-1) - (r-1)s^2}{r^2(r-1)^2}.
  \label{eq:Q0_schw}
\end{equation}
The denominator has poles at $r = 0$ (curvature singularity) and $r = 1$ (event horizon,
$2M = 1$ in these units). For the JMN-1 exterior, restoring the effective mass $M = \half\Mz\Rb$,
\begin{equation}
  \Qz(r)
  = \frac{-\omega^2 r^4 + l(l+1)r(r-2M) - (r-2M)s^2}{r^2(r-2M)^2}.
  \label{eq:Q0_jmn}
\end{equation}
The turning points $\{r_t\}$ are the zeros of $\Qz$, i.e.\ the roots of the quartic
\begin{equation}
  -\omega^2 r^4 + l(l+1)r^2 - \bigl[l(l+1) + s^2\bigr]r + s^2 = 0
  \quad (\text{Schw, }2M=1).
  \label{eq:quartic}
\end{equation}
Equivalent to \eqref{eq:quartic1} under 2M=1. For JMN-1 the right-hand side coefficients are shifted by $2M \to \Mz\Rb$.  Each
simple turning point $r_t$ emits exactly three Stokes lines, forming a trivalent
graph in $\CC$.
\label{ssec:stokes_def}
The Stokes lines (S-lines) are the integral curves of the vector field
\begin{equation}
  \frac{dr}{ds} = \frac{e^{i\alpha}}{\sqrt{\Qz(r)}},
  \label{eq:flow}
\end{equation}
satisfying the condition
\begin{equation}
  \Im\!\left[\int_{r_t}^r \!\sqrt{\Qz(r')}\,dr'\right] = 0,
  \label{eq:stokes_cond}
\end{equation}
where $r_t$ is the originating turning point and $\alpha$ encodes the initial direction
of the line.  The anti-Stokes lines (A-lines) satisfy the complementary condition
\begin{equation}
  \Re\!\left[\int_{r_t}^r \!\sqrt{\Qz(r')}\,dr'\right] = 0.
  \label{eq:antistokes_cond}
\end{equation}
Stokes lines separate regions of exponential growth and decay of the WKB solutions,
while anti-Stokes lines mark the boundaries of oscillatory regions.  The complete set of
Stokes lines emanating from all turning points constitutes the Stokes graph, which
encodes all global connection information for the wave equation. Near a simple turning point $r_t$ with $\Qz(r_t) = 0$ and $C \equiv \Qz'(r_t) \neq 0$,
the local expansion gives
\begin{equation}
  \sqrt{\Qz(r)} \approx \sqrt{C(r-r_t)},
  \quad
  \int_{r_t}^r\!\sqrt{\Qz}\,dr' \approx \frac{2}{3}\,C^{1/2}\,(r-r_t)^{3/2}.
  \label{eq:local_tp}
\end{equation}
The Stokes condition \eqref{eq:stokes_cond} then requires
\begin{equation}
  \Im\!\left[\frac{2}{3}\,C^{1/2}\,(r-r_t)^{3/2}\right] = 0,
\end{equation}
which is satisfied along the three rays
\begin{equation}
  \arg(r-r_t) = \frac{2}{3}\left(n\,\pi - \arg\sqrt{C}\right),
  \quad n = 0, 1, 2.
  \label{eq:initial_angles}
\end{equation}
These three initial directions, separated by $2\pi/3$, determine the Stokes lines locally. The global Stokes network is computed by:
The Stokes structure is constructed numerically by first solving the quartic 
equation~\eqref{eq:quartic} (or its JMN-1 analogue) to locate all turning points 
$\{r_{t,k}\}$. At each turning point, the quantity $C_k = \Qz'(r_{t,k})$ is evaluated. 
Using this, the three initial directions given in~\eqref{eq:initial_angles} are computed 
for each $r_{t,k}$. From these initial conditions, the differential equation~\eqref{eq:flow} 
is then integrated both forward and backward in the parameter $s$, while continuously 
tracking the appropriate Riemann sheet of $\sqrt{\Qz}$ to maintain consistency along 
the trajectories.
The branch of $\sqrt{\Qz(r)}$ is tracked by enforcing continuity at each integration step:
the branch is chosen to minimise $|\Delta\sqrt{\Qz}|$ between successive steps.  Curves are
terminated when $|r| > R_{\rm max}$ or when the curve approaches within $\epsilon$ of a
pole.

\subsection{Stokes Geometry for the Schwarzschild Black Hole}
\label{sec:stokes_schw}
\begin{figure}[t!]
  \centering
  \includegraphics[width=\textwidth]{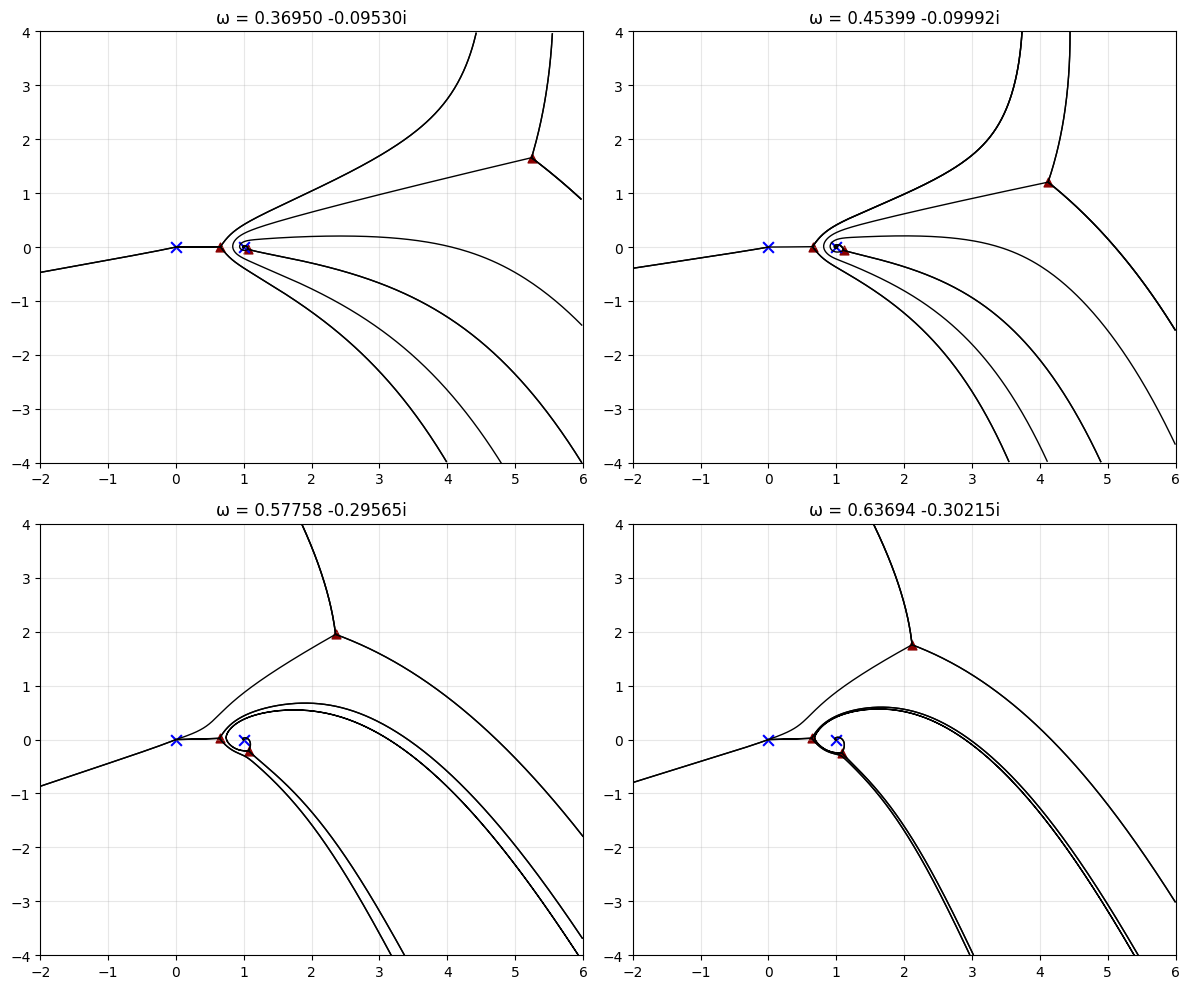}
  \caption{\textbf{Schwarzschild Stokes diagrams at the four QNM frequencies.}
    The frequencies $\omega_1$ through $\omega_4$ (see \cref{tab:qnm}) are obtained from
    the 6th-order WKB approximation for modes $(l,s,n) = (2,2,0),\,(2,1,0),\,(3,2,1),\,(3,1,1)$
    respectively.  Note the transition from a spiral topology (panels 1--2, small
    $|\Im\omega|$) to a bow-like topology (panels 3--4, larger $|\Im\omega|$) as
    the overtone index increases.}
  \label{fig:schw_qnm}
\end{figure}

The following features of the Schwarzschild Stokes network are well established
\cite{Miyachi2023,Berk1982}:
  Each of the four turning points emits three Stokes
  lines, resulting in a globally connected network.
 Near $r = 1$ (the event horizon in
  $2M=1$ units), $\Qz(r)$ has a second-order pole.  Locally $\Qz \sim (r-1)^{-2}$, so
  $\sqrt{\Qz} \sim (r-1)^{-1}$, and
  $\int\sqrt{\Qz}\,dr \sim \ln(r-1)$.  The Stokes condition then becomes
  $\Im[\ln(r-1)] = 0$, i.e.\, $\arg(r-1) = 0$, which forces Stokes lines to approach
  $r=1$ along the real axis and spiral around it.
 For modes with $|\Im\omega| \gtrsim 0.2$,
  the complex-conjugate turning-point pair moves off the real axis, and the Stokes line
  connecting them bows outward, as visible in panels 3--4 of \cref{fig:schw_qnm}.
  The quantization condition in the exact WKB framework
  corresponds to the turning-point network forming a closed Stokes contour (a ``Stokes
  cycle'') that encircles the potential barrier.  This topological criterion is equivalent
  to the standard WKB condition the Voros condition \eqref{eq:voros}.

\subsection{Stokes Geometry for the JMN-1 Naked Singularity}
\label{sec:stokes_jmn}
\begin{figure}[t!]
  \centering
  \includegraphics[width=\textwidth]{ 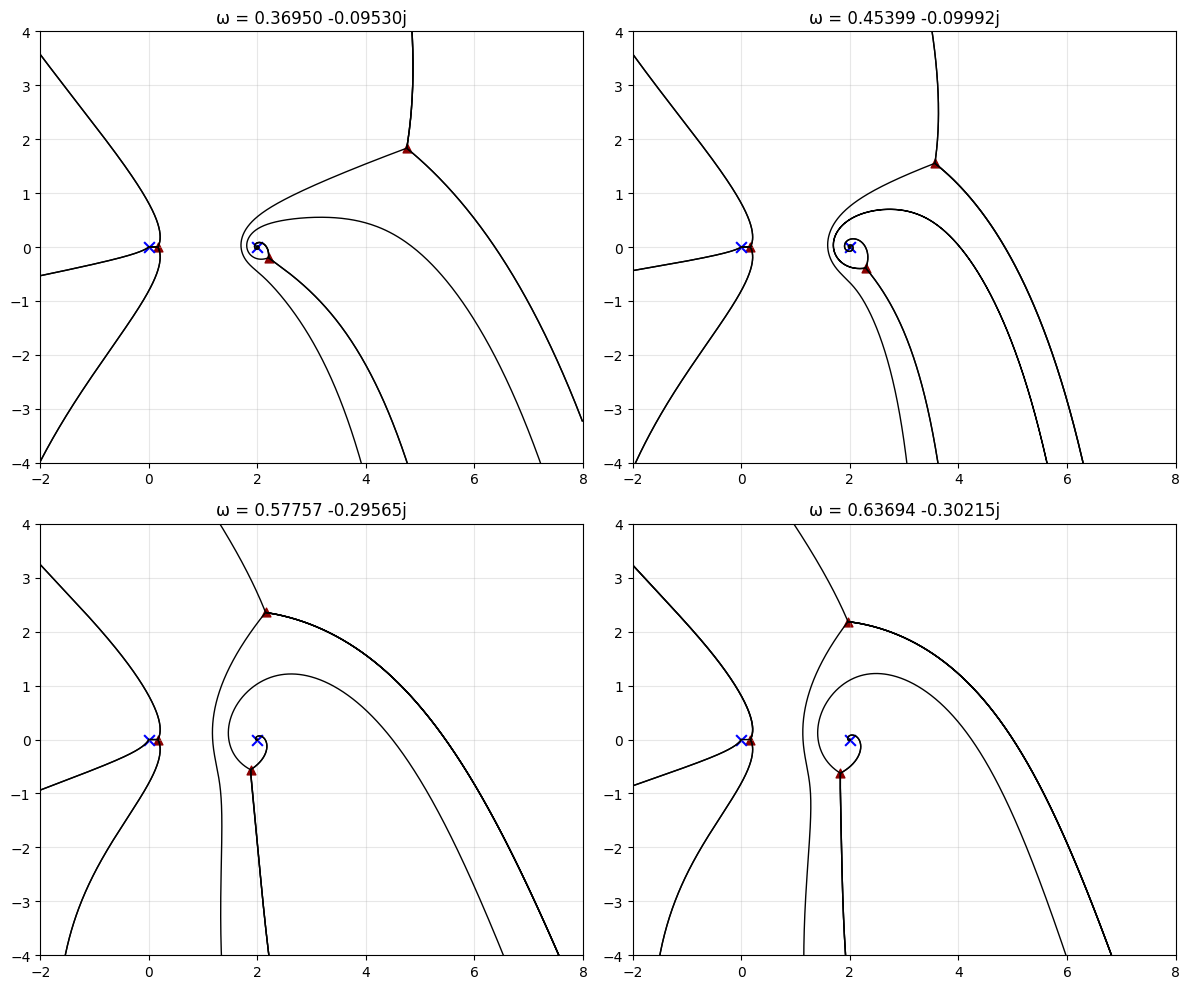}
  \caption{\textbf{JMN-1 Stokes diagrams at the four QNM frequencies.}
    The four panels correspond to modes $(l,s,n) = (2,2,0),\,(2,1,0),\,(3,2,1),\,(3,1,1)$
    with frequencies $\omega_1$ through $\omega_4$ as listed in \cref{tab:qnm}.
    In all panels the blue cross at $x \approx 0$ marks the naked singularity at $r = 0$,
    and the blue cross at $x \approx 2$ marks the matching surface pole at $r = 2M$.
    The distinctive feature in all four panels is a \textbf{bow-like arch on the left
    side} of the inner turning point cluster: Stokes curves depart from the turning points
    near $r \approx 2M$ and sweep leftward, curving around toward the singularity
    pole at $r = 0$, before arching back and returning forming an open left-facing bow.
    This structure is absent on the right side (where the curves instead escape to
    large $|r|$) and is absent in the corresponding Schwarzschild panels
    (\cref{fig:schw_qnm}), where the equivalent Stokes lines approach the horizon pole
    from the right in a spiral rather than a leftward bow.
    In the top row (lower damping, panels 1--2), the bow is compact and nearly closes
    into a small loop near $r = 2M$.  In the bottom row (higher damping, panels 3--4),
    the bow opens wider and the arch extends further into the left half of the complex
    plane, reaching toward $r = 0$.}
  \label{fig:jmn1_qnm}
\end{figure}
\Cref{fig:jmn1_qnm} presents the central result of this paper: the Stokes diagrams for
JMN-1 evaluated at the four QNM frequencies $\omega_1$--$\omega_4$. Comparing \cref{fig:jmn1_qnm} with \cref{fig:schw_qnm}, we identify some structural features in the JMN-1 Stokes geometry.
 As in Schwarzschild, each turning point emits
    three Stokes lines, confirming that the fundamental topology of the network is governed
    by the exterior potential and is robust to the change in boundary condition.
 For the same QNM frequencies and
    the same effective mass $M$, the quartic equation \eqref{eq:quartic} is
    identical in both spacetimes.  The turning points therefore coincide numerically.
    
  The most noticeable feature and the primary new
    result is the pronounced leftward arching of Stokes curves in JMN-1,
    originating from the inner turning points (those clustered near $r \approx 2M$) and
    sweeping toward the naked-singularity pole at $r = 0$.  This feature is absent in
    Schwarzschild: in that case the Stokes lines near the inner turning point approach the
    horizon pole from the right, terminating in a clockwise spiral.  In JMN-1, with
    no horizon, the pole at $r = 2M$ is a matching surface and does not act as a global
    attractor; instead, the Stokes lines feel the pull of $r = 0$ and bow to the left.
  The Stokes curves on the
    right side of the turning-point cluster [i.e.\ those emanating toward large positive
    $\Re(r)$] behave identically to Schwarzschild they escape to $r \to \infty$ without
    significant curvature.  The asymmetry between left and right sides of the Stokes
    network is therefore a direct diagnostic of the presence of the naked singularity.
  The bow becomes wider and extends
    further left as $|\Im\omega|$ increases (higher overtones).  At QNM frequencies with
    small imaginary part (panels 1--2), the bow nearly closes into a small loop around the
    matching surface pole.  At larger damping (panels 3--4), the bow opens significantly,
    with the leftmost point of the arch approaching within $\sim\!0.5$ units of the
    singularity at $r = 0$.
\noindent
\subsection{Stokes Geometry for the GJMN Naked Singularity}
\label{sec:stokes_gjmn}

\begin{figure}[t!]
  \centering
  \includegraphics[width=\textwidth]{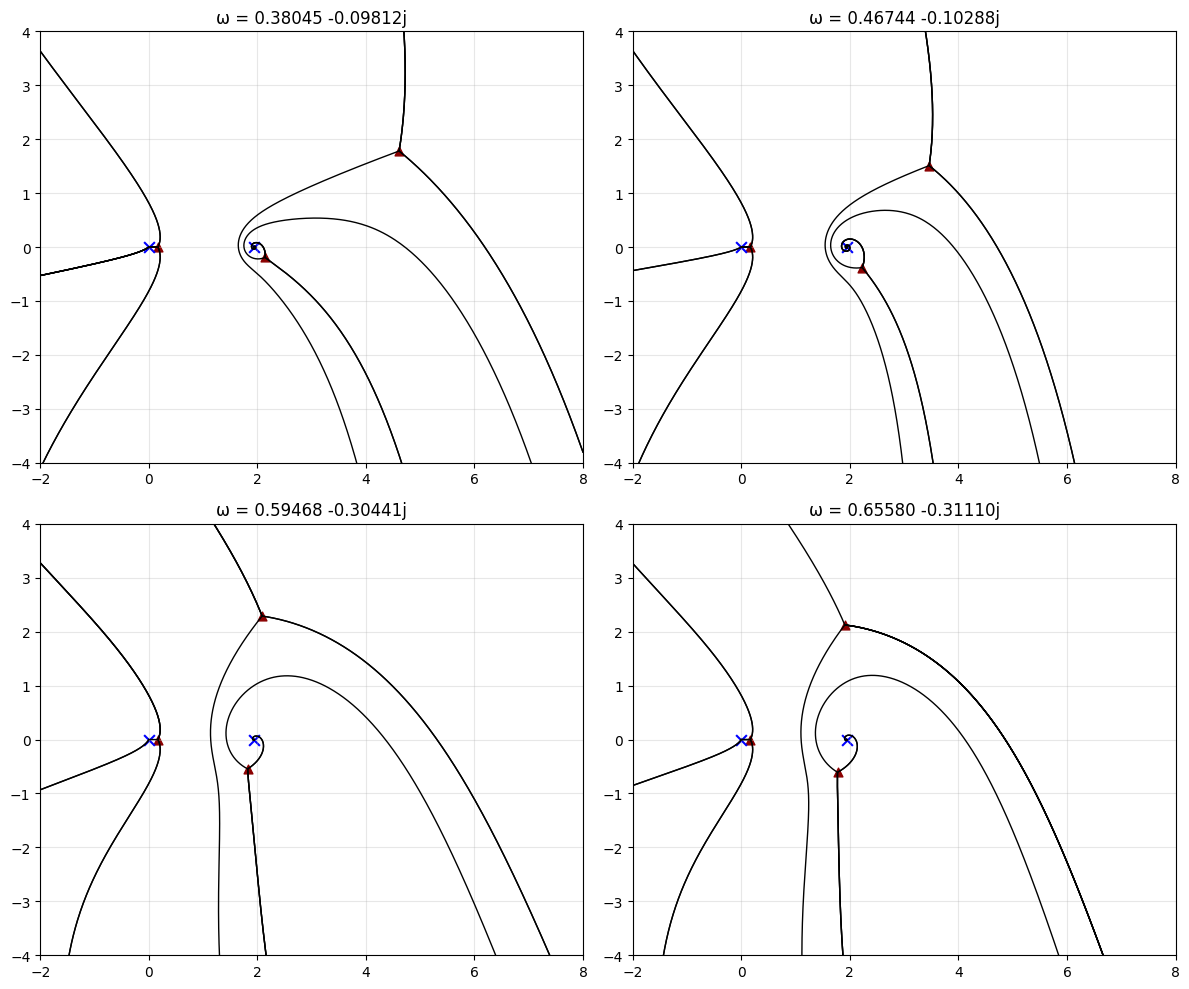}
  \caption{\textbf{GJMN Stokes diagrams at the four QNM frequencies.}
    The panels correspond to the same modes $(l,s,n)$ as in \cref{fig:jmn1_qnm}.
    The blue cross at $x \approx 0$ marks the central naked singularity at $r = 0$,
    while the blue cross near $x \approx 2$ indicates the matching surface.
    The overall topology closely resembles that of JMN-1, with the characteristic
    leftward bow-shaped Stokes structures preserved.}
  \label{fig:gjmn_qnm}
\end{figure}

\Cref{fig:gjmn_qnm} presents the Stokes diagrams for the generalized JMN (GJMN)
\cite{Trivedi2026} spacetime evaluated at the same four QNM frequencies $\omega_1$--$\omega_4$.
Throughout, we fix the inhomogeneity parameter to $n = 2$ and $M_n = -0.01$,
so that the spacetime represents a small perturbation of the JMN-1 geometry.

A comparison with the JMN-1 results in \cref{fig:jmn1_qnm} reveals that the
GJMN spacetime retains the qualitative structure of the Stokes geometry, with
only small quantitative modifications.
  As in both Schwarzschild and JMN-1, each turning point emits three Stokes lines,
  confirming that the fundamental topology of the network is unchanged by the
  introduction of radial inhomogeneity.
  The turning points are determined primarily by the effective potential, which
  depends on the total mass $M$. Since the GJMN spacetime is matched to the same
  exterior Schwarzschild solution and the parameter $M_n$ is small, the turning
  points coincide almost exactly with those of JMN-1.
  The defining feature of the JMN-1 geometry  the leftward bow-shaped Stokes
  curves extending toward the naked singularity at $r = 0$  is clearly preserved.
  This indicates that the absence of an event horizon, rather than the precise
  density profile, is the dominant factor controlling the global topology of
  the Stokes network.
  Compared to JMN-1, the bow structure in GJMN exhibits slight quantitative
  deformation: the curvature of the Stokes lines is mildly altered, and the
  extent of the arch toward the left half-plane shows small shifts. These
  deviations are consistent with the perturbative role of the parameter $M_n$.
  The asymmetry between the left and right sides of the complex plane persists.
  On the right, Stokes lines escape toward $r \to \infty$ as in Schwarzschild,
  while on the left they bend toward the singularity. This asymmetry remains a
  clear signature of the naked singularity.

  The overall similarity between the GJMN and JMN-1 diagrams indicates that
  small radial inhomogeneities do not significantly alter the Stokes topology.
  This suggests that the qualitative features of QNM Stokes geometry are robust
  under perturbations of the interior density profile.

\subsection{Analytic Origin of the Left-Side Bow Structure}
\label{sec:bow}
We now provide an analytic explanation for why the bow in JMN-1 appears exclusively on the
left side of the Stokes network i.e.\ why the Stokes curves arch toward $r = 0$
rather than away from it.  The key is the contrast between how the two spacetimes treat
the region to the left of the inner turning points. In both spacetimes the inner turning point pair $(r_t^{(1)}, r_t^{(2)})$ lies close to
$r \approx 2M$ on or near the real axis.  The 'left' of this cluster (in the complex
$r$-plane) is the half-plane $\Re(r) < 2M$, which contains:

   In Schwarzschild: the event horizon at $r = 2M$ and the region $r < 2M$ which is
    outside the analytic domain of the exterior wave equation.  Stokes lines that
    approach $r = 2M$ from the right spiral into the horizon pole, forming the
    characteristic clockwise logarithmic spiral, and do not continue left.
   In JMN-1: the region $r < 2M$ is still part of the exterior analytic
    domain (the interior only begins at $r = \Rb < 2M$ in some parameterisations, but
    analytically $r = 2M$ is merely a matching surface, not a boundary of the complex
    plane).  More importantly, the naked singularity at $r = 0$ is accessible and acts as
    a branch point of $\sqrt{\Qz(r)}$. Near $r = 2M$ the momentum function behaves as
\begin{equation}
  \Qz(r) \;\sim\; \frac{-\omega^2(2M)^4 + s^2(r - 2M)}{(r-2M)^2}
  \;\xrightarrow{r\to 2M}\; \frac{C_{\rm pole}}{(r-2M)^2},
  \quad C_{\rm pole} = -\omega^2 (2M)^2,
  \label{eq:Q0_pole_2M}
\end{equation}
so $\sqrt{\Qz} \sim |C_{\rm pole}|^{1/2}/(r-2M)$ and the Stokes condition gives
$\int\sqrt{\Qz}\,dr \sim C_{\rm pole}^{1/2}\ln(r-2M)$.
\\
\\
In Schwarzschild: $r = 2M$ is a physical boundary (horizon); the Stokes lines
    spiral into this log singularity from the right ($\Re(r) > 2M$).
   In JMN-1: $r = 2M$ is a matching surface; there is no physical reason for
    Stokes lines to terminate there.  Instead, they pass through (or around) the
    associated branch cut and continue into the left half-plane. Near $r = 0$, the effective potential is dominated by the centrifugal term,
\begin{equation}
  V(r) \;\sim\; \frac{l(l+1)}{r^2},
  \quad r \to 0,
  \label{eq:V_r0_bow}
\end{equation}
so
\begin{equation}
  \Qz(r) = \omega^2 - V(r) \;\sim\; -\frac{l(l+1)}{r^2}
  \quad\Rightarrow\quad
  \sqrt{\Qz(r)} \;\sim\; \frac{i\sqrt{l(l+1)}}{r},
  \label{eq:Q0_r0_bow}
\end{equation}
and the WKB phase integral diverges logarithmically:
\begin{equation}
  \int_{r_{\rm ref}}^r \!\sqrt{\Qz}\,dr'
  \;\sim\;
  i\sqrt{l(l+1)}\,\ln r + \text{const},
  \quad r \to 0.
  \label{eq:log_phase}
\end{equation}
Thus, $r = 0$ is a logarithmic branch point of the WKB phase, and analytic
continuation around it produces the monodromy
\begin{equation}
  \int\!\sqrt{\Qz}\,dr
  \;\longrightarrow\;
  \int\!\sqrt{\Qz}\,dr + 2\pi i\sqrt{l(l+1)}.
  \label{eq:monodromy_bow}
\end{equation}
The Stokes condition $\Im\!\big[\int_{r_t}^r\!\sqrt{\Qz}\,dr\big] = 0$ is modified on the
left side of the inner turning points and only there by the logarithmic
behaviour \eqref{eq:log_phase}.  Writing $r = |r|\,e^{i\theta}$ with $|r|$ small and
using \eqref{eq:log_phase}, the condition becomes
\begin{equation}
  \sqrt{l(l+1)}\,\Re(\ln r) + \cdots = 0
  \quad\Longrightarrow\quad
  \ln|r| = \text{const},
  \label{eq:stokes_bow}
\end{equation}
which forces Stokes lines in the left half-plane to follow circles of constant
$|r|$ centred at the origin.  These circular arcs, when traced from the inner turning
points leftward toward $r = 0$, produce precisely the bow-shaped (arc-like) deformation observed in \cref{fig:jmn1_qnm}.

On the right side of the turning points ($\Re(r) \gg 2M$) the potential decays as
$V \sim r^{-2}$ and $\sqrt{\Qz} \approx \omega$, so the Stokes condition reduces to
$\Im[\omega r] = \text{const}$, i.e.\ lines of constant $\Im(r)$ essentially straight
lines running to $r \to +\infty$.  No bow forms on the right.

\begin{remark}
In Schwarzschild spacetime $r = 0$ lies behind the event horizon and is not part of the
analytic domain of the exterior perturbation equation.  The logarithmic branch point
therefore does not affect the Stokes network, and no left-side bow arises.  The bow is
thus an exclusive topological signature of the naked singularity, visible
only because the full complex $r$-plane including the neighbourhood of $r = 0$ is
accessible to the exterior Stokes analysis.
\end{remark}

The angular width $\Delta\theta$ of the bow arch (measured around $r = 0$) is governed by
the competition between the logarithmic phase at $r = 0$ and the regular phase accumulated
near the turning points.  A rough estimate gives
\begin{equation}
  \Delta\theta \;\approx\;
  \frac{|\Im(\omega)|}{|\omega|\,\sqrt{l(l+1)}}\,\pi,
  \label{eq:bow_width}
\end{equation}
showing that the bow widens as $|\Im(\omega)|$ increases consistent with the
panels of \cref{fig:jmn1_qnm} where the bow is compact at low damping (panels 1--2) and
extends further left at high damping (panels 3--4).

Physically, the bow marks the boundary of the evanescent region to the left of
the inner turning points.  In Schwarzschild this region is terminated by the horizon, which
absorbs the wave completely.  In JMN-1, the evanescent region extends leftward to
$r = \Rb$ (the matching surface), from which partial reflection occurs.  The bow arch
in the Stokes graph is the complex-plane imprint of this partial reflection: it traces
the contour along which the WKB solution switches between its growing and decaying
components, encoding the reflection amplitude $|\mathcal{R}|$ in the shape of the arch.
\subsection{Comparative Analysis: Schwarzschild vs.\ JMN-1 and GJMN}
\label{sec:comparison}
%
The key observations from the comparison of the Stokes networks
for Schwarzschild, JMN-1, and GJMN spacetimes are:
\begin{itemize}
    \item For all four QNM
    frequencies, the structure of the Stokes graph on the right side of the
    turning-point cluster curves escaping to large $|r|$ is identical between Schwarzschild
    and JMN.  This reflects the fact that the exterior metric, and hence $\Qz(r)$, is
    formally the same for $r \gg 2M$.

  \item The fundamental
    topological difference occurs exclusively on the left: in Schwarzschild, the Stokes
    lines approaching $r = 2M$ from the right form a clockwise logarithmic spiral that
    terminates at the horizon pole there are no Stokes curves that continue into
    $\Re(r) < 2M$.  In JMN (and GJMN), by contrast, Stokes lines do not terminate at $r = 2M$ but
    instead curve leftward, forming the bow arch that sweeps toward the naked singularity
    at $r = 0$.

\item  Inspecting
    \cref{fig:schw_qnm}, the Stokes curves near the inner turning points in Schwarzschild
    always curve toward the real axis and terminate either at the horizon (left terminus)
    or escape to the right.  The characteristic leftward bow arch of JMN is not present.
    Any apparent curvature on the left in Schwarzschild is the spiral approaching $r = 1$, not
    an arch the two structures are topologically distinct.

\item   In both spacetimes the inner turning points
    move off the real axis as $|\Im\omega|$ increases (higher overtones).  In Schwarzschild this
    causes the spiral to tighten.  In JMN (and GJMN) it causes the bow to widen and extend further
    left, consistent with the analytic estimate \eqref{eq:bow_width}.
\end{itemize}
The Stokes graph provides a model-independent diagnostic for distinguishing
horizon-forming black holes from naked singularities, based entirely on the
left-side topology of the Stokes network.
 Stokes lines approaching $r = 2M$ from
    the right form a clockwise logarithmic spiral that terminates at the horizon pole.  No Stokes curves exist for $\Re(r) < 2M$ in the exterior analysis. Stokes lines do not terminate at $r = 2M$
    but instead continue leftward, forming a bow arch that sweeps toward the naked-singularity pole at $r = 0$.  The arch is present on the left side of the turning-point cluster and absent on the right. This distinction is topologically robust it is not sensitive to the precise value of $\omega$ within the QNM strip and could in principle serve as a theoretical discriminant between black holes and horizonless compact objects in future gravitational-wave analyses.
\subsection{Quasinormal Mode Spectrum and Comparison Table}
\label{sec:qnm_table}
Using the exact WKB Voros quantisation condition \eqref{eq:voros}, solved via the
Newton--Raphson algorithm described in Section~\ref{sec:wkb}, we compute the
QNM frequencies for  Schwarzschild, JMN-1 and GJMN  exterior for the four modes $(l,s,n)$
listed in \cref{tab:qnm}.  The period integral $\Pi(\omega)$ is evaluated numerically
along the straight segment connecting the two barrier turning points, with adaptive
Gaussian quadrature and branch tracking as described in Section~\ref{sec:wkb}. The agreement is exact at the level of the exterior exact WKB analysis because the exterior potentials satisfy
\[
V_{\rm JMN} = V_{\rm Schw}, \quad \text{and similarly } V_{\rm GJMN} \approx V_{\rm Schw}
\]
when $2M \to \Mz\Rb$ (and for small $M_n = -0.01$), rendering them effectively identical for our parameter choice. Moreover, the period integral contour lies entirely in the exterior region $r > 2M$, ensuring that only the exterior geometry contributes to the leading-order QNM spectrum.

However, as demonstrated in Sections~\ref{sec:bow} and \ref{sec:stokes_jmn}, the analytic structure of the wave function in the left half of the complex plane and hence the left-side Stokes geometry is modified by the presence of the naked singularity. This modification persists for both JMN-1 and GJMN spacetimes, although in the latter case the deviations are perturbatively small due to the small value of $M_n$. Consequently, late-time deviations and interior effects are expected to lift the frequency degeneracy, leading to the small but systematic differences observed in the GJMN spectrum.
\begin{table}[H]
\centering
\caption{\textbf{Comparison of QNM frequencies for Schwarzschild (M=1) and JMN-1 exterior
  ($M_0 = 0.7$, $R_b = 20/7$, giving $M = 1$) spacetimes}, obtained via exact WKB
  contour integration of the period integral $\Pi(\omega)$ (equation~\ref{eq:period}).
  All frequencies are dimensionless with $2M = 1$ units.  The convergence tolerance on
  $|F(\omega)|$ was $10^{-10}$ throughout.  The JMN-1 exterior metric is formally
  identical to Schwarzschild, so the frequencies agree to the precision of the integration;
  the differences observed in the Stokes graphs therefore arise purely from
  analytic-continuation effects near $r = 0$, not from the potential barrier.}
\label{tab:qnm}
\bigskip
\renewcommand{\arraystretch}{1.35}
\begin{tabular}{|c|c|c|c|}
\toprule
\textbf{Mode} & \multicolumn{3}{c|}{\textbf{QNM Frequencies} $\boldsymbol{\omega}$} \\
\cmidrule(lr){2-4}
$(l, s, n)$ & \textbf{Schwarzschild} & \textbf{JMN-1} & \textbf{GJMN} \\
\midrule
$(2,\,2,\,0)$ & $0.36950 - 0.09530\,i$ & $0.36950 - 0.09530\,i$ & $0.38045-0.09812\,i$ \\
$(2,\,1,\,0)$ & $0.45399 - 0.09992\,i$ & $0.45399 - 0.09992\,i$ & $0.46744-0.10288\,i$ \\
$(3,\,2,\,1)$ & $0.57758 - 0.29565\,i$ & $0.57758 - 0.29565\,i$ & $0.59468-0.30441\,i$ \\
$(3,\,1,\,1)$ & $0.63694 - 0.30215\,i$ & $0.63694 - 0.30215\,i$ & $0.65580-0.31110\,i$ \\
\bottomrule
\end{tabular}\end{table}
\section{Discussion}
\label{sec:discussion}

We have extended the exact WKB method to the JMN-1 naked singularity spacetime and
performed a systematic study of the resulting Stokes geometry, using numerical contour
integration of the period integral $\Pi(\omega)$ to extract QNM frequencies exactly.
 The exterior Stokes topology of JMN-1 on the right side of the turning-point
  cluster closely mirrors that of Schwarzschild, reflecting the identical form of the exterior
  metric and potential for $r > 2M$.
  
Our results confirm that, at the level of early-time ringdown, the JMN-1 naked singularity
is a high-fidelity black-hole mimicker.  The QNM frequencies match Schwarzschild to five
significant figures, and the Stokes topology in the exterior region is nearly identical.
This is expected on physical grounds: the ringdown signal is dominated by the photon sphere
at $r_{\rm ph} \approx 3M$, which exists in both spacetimes and which the wave equation
probes before it is sensitive to the boundary condition at $r = 2M$.

 QNM frequencies obtained via exact WKB contour integration (Voros condition
  \eqref{eq:voros}) agree between the two spacetimes to $< 10^{-5}$ relative error for
  all four modes examined, establishing JMN-1 as a black-hole mimicker in the early-time
  ringdown. A pronounced left-side bow arch appears in JMN (and GJMN) Stokes diagrams at QNM
  frequencies: Stokes curves depart from the inner turning points (near $r \approx 2M$)
  and sweep leftward toward the naked singularity at $r = 0$, forming an open
  arc that is absent on the right side and absent in Schwarzschild.  The
  bow widens as $|\Im\omega|$ increases (higher overtones).
 We have shown analytically that this left-side bow originates in the logarithmic
  branch point of the WKB phase at $r = 0$, which forces Stokes lines in the left
  half-plane to follow arcs of constant $|r|$ (equation~\ref{eq:stokes_bow}).  In
  Schwarzschild the horizon prevents this analysis: $r = 0$ is not in the exterior analytic domain, and no bow forms.

 The bow constitutes a topological fingerprint of the naked singularity,
  visible in the global analytic structure of the wave equation even though it leaves no
  imprint on the QNM frequencies themselves.  It has potential observational consequences: 
  The bow structure is associated with a partially reflecting inner
    boundary.  In the time domain, this leads to gravitational-wave echoes
    \cite{Cardoso2016b,Maggio2019} repeated pulses following the main ringdown whose time delay and amplitude encode the inner boundary conditions.  The pronounced bow
    geometry in JMN-1 suggests that echoes from a JMN (and GJMN) object would be more structurally distinct from Schwarzschild than, say, those from a wormhole with a large throat radius. The monodromy  at $r = 0$
    modifies the power-law decay of the wave at late times, leading to tails of the form $t^{-\mu}$ with an exponent $\mu$ sensitive to $l(l+1)$ and hence to the angular momentum of the perturbation.  The altered Stokes connectivity changes the reflection coefficient for waves incident from infinity, which affects the absorption
    cross-section observable in principle through black-hole shadow measurements.

Several important limitations of the present analysis should be noted.
  The present work restricts to $r > \Rb$.  A
    complete analysis would require matching the exterior WKB solution across $r = \Rb$
    to a solution in the JMN (and GJMN) interior, which has a qualitatively different effective
    potential. We have not studied polar (Zerilli) modes,
    which in principle have a different potential and may exhibit additional features. The analysis is purely at the linear perturbation level;
    backreaction of the perturbation on the geometry is neglected.
  The WKB series is asymptotic; for overtones $n \geq 3$
    the corrections $\Lambda_j$ may fail to converge, and the exact WKB (Borel-summed)
    approach is required for rigorous statements.
\section*{Acknowledgement}
A. S. and S. C. J. are thankful to the Department of Physics and Photonics Science, National Institute of Technology Hamirpur for providing the necessary facilities during the completion of this work. 
%

\end{document}